\documentclass[%
 aip,
 amsmath,amssymb,
 print,%
]{revtex4-1}

\usepackage{graphicx}
\usepackage{dcolumn}
\usepackage{bm}

\begin{document}


\title{A multi-parameter Lie transform method applied to the transform of the Lagrangian differential one-form}

\author{Shuangxi Zhang}
\email{zshuangxi@gmail.com}
\affiliation{Graduate School of Energy Science, Kyoto University, Uji, Kyoto 611-0011, Japan.}


\date{\today}

\begin{abstract}
In this paper, to fit the multi-scale perturbations, the single-parameter Lie transform perturbation theory given in [Ann Phys, 151 1 (1983)] is generalized to a multi-parameter case, which provides a formal solution of the new Lagrangian differential 1-form transformed from the old one. In the new method, the generators and their orders are appropriately chosen under the purpose of reducing the fast variables from the new Lagrangian differential 1-form order by order. As for the application, this multi-parameter Lie transform method is applied to deriving the gyrokinetic model with the presentation of low-frequency electrostatic perturbations. Compared with the conventional model, the Finite Larmor Radius terms in the Lagrangian 1-form of the new model are perfectly cancelled, without using any gauge function.
\end{abstract}

\maketitle

\section{Introduction}\label{intro}

The perturbations exist commonly in the classical mechanical systems, e.g., the planet's elliptical orbit surrounding the star is perturbed by the weak gravitational force generated by other neighboring planets \cite{1989arnoldbook,moser1978}; the guiding-center orbit of charged particles moving in the strong magnetic field suffers the perturbation of the gyromotion as well as electromagnetic perturbations \cite{1990brizard,1967hastie, 1978northrop}. When  perturbations exist, the simple structures or the symmetries of the original system in the old coordinates, i.e, the pair of action and angle, are broken that for some cases, it  becomes hard to solve the system in the old coordinates. A theoretical method to overcome the difficulty in the old coordinate frame is to find a new coordinate frame through a coordinate transform. In the new coordinate frame, the fast variables are reduced from the whole system up to some order of the parameters \cite{1989arnoldbook, 1947kryloff}. Then, the simple structure or the symmetries may be recovered in the new coordinate systems.

For the canonical systems, the generating function method for the coordinate transform is widely adopted. This method is based on Hamilton-Jacobi equations \cite{abudullaev2006, 1989arnoldbook, 2006marsden}. Another approach for the canonical system is to carry out Lie transform over the equations of motion \cite{1966hori,deprit1969}. There is a third transform method, which imposes a pullback transform on the Lagrangian differential 1-form. This method was firstly introduced by Littlejohn and Cary \cite{1982littlejohn, 1983cary} and is widely adopted in the research of  gyrokinetic theory \cite{1988hahm, 1990brizard, 2007brizard1, 2009cary, 2010scott, 2009miyato, 2013burby, mcmillan2014, 2016tronko, 2000qinhong1, 1993hammett, 2000sugama, 2010garbet, 2006shaojiewang, 2000jenko, 1993parker}. The focus of this paper is on this method. In the original literature \cite{1983cary}, this method is specially designed for noncanonical systems.

As it's well-known, implementing the extremal principle to the action as the integral of the Lagrangian over the time could lead to the equations of motion. If fast variables are reduced from the dynamical system, new equations of motion can be derived immune from the fast variables. The purpose of the application of Lie transform perturbation theory (LTPT) to the Lagrangian differential 1-form in terms of perturbations, is to obtain a new 1-form with the fast variables reduced from the dynamical system up to some order. Just as the derivation in Ref.(\onlinecite{1983cary}) shows, after expanding the formal formula of the new 1-form by the order of the small parameter, the terms depending on the fast variables are cancelled at each order of the new 1-form by the properly choosing generators. Therefore, one important step is to obtain the formal formula of the new 1-form defined on the new coordinate frame.

In Ref.(\onlinecite{1983cary}), the single-parameter LTPT is generalized to the high order situation with the perturbive differential 1-form in the old coordinate frame expressed by the formula $\gamma  = {\gamma _0} + \varepsilon {\gamma _1} + {\varepsilon ^2}{\gamma _2} +  \cdots$. The formula of the new one  in the new coordinate frame transformed from the old one is given as $\Gamma  = {T_n} \cdots {T_2}{T_1}\gamma ({\bf{Z}}) + dS$ with ${T_i} = \exp \left( { - {\varepsilon ^i}{L_{{{\bf{g}}_i}}}} \right)$. The operator $L_{\mathbf{g}_i}$ will be illustrated in Sec.(\ref{sec0}). However, for some physical problems, the perturbation included by the Lagrangian 1-form is much more complex that a sequence of ordering parameter, like $(\varepsilon,\varepsilon^2,\cdots)$, is not enough.  For example, in the magnetized plasma system, there exists in fact multiple scales contained by the perturbations \cite{1980antonsen, 1982frieman, 2009fujisawa1, 2011bottino, 2016chenliu, 2010diamond1}, each of which may be characterized by a different parameter. The first one in the magnetized plasma is the Larmor radius, which is much smaller than the spatial characteristic length of the equilibrium magnetic field. The second is the electrostatic and electromagnetic perturbations, the amplitudes of which may be different from each other. The quantities of spatial gradient length and oscillating frequency of the perturbations usually can't be treated as of order $O(1)$. Although there may exist a physical mechanism to associate those quantities with each other, we actually don't know how each parameter depends on others before we eventually obtain their values or formulas to quantitatively give the relation between them.  Based on these facts, it may be more useful to treat each parameter as an independent one from others.  If so, the perturbation theory applied to those problems and depending on multiple parameters needs also to be generalized to include multiple parameters.

This paper is dedicated to developing a Lie transform method including multiple perturbative parameters, which is specially applied to the pullback transform of the differential 1-form for the reduction the fast variables from the dynamical system. The whole scheme begins with generalizing the single-parameter LTPT \cite{1983cary} to the kind including multiple parameters with the new 1-form given by an analytical formula $[\exp(-\mathbf{E}_n\cdot L_{\mathbf{g}})\gamma](\bf{Z})$, where $\mathbf{E}_n\equiv (\varepsilon_1,\cdots,\varepsilon_n)$ is the parameter vector and $\mathbf{g}=(\mathbf{g}_1,\cdots,\mathbf{g}_n)$ is the generator vector, and $\bf{Z}$ is the new coordinate. The limitation of the formal formula is that all the parameters in $\mathbf{E}_n$ are independent of each other, which is not practical for realistic application. To overcome this defect, we formally relax the limitation over the parameter sequence, so that all parameters needn't to be independent of each other as described in Subsec.(\ref{sec3.1}). This relaxation of the limitation on the parameters has special superiority that the parameters can be freely chosen to satisfy the cancellation operation for the new 1-form explained in Subsec.(\ref{sec3.2}). In fact, for the complex perturbations, such as those in  magnetized plasmas as explained before, several perturbative scales coexist, so that we usually can't determine what perturbative parameter should be introduced to the formal formula of new 1-form before carrying out the cancelling procedure. This is different from the single parameter theory \cite{1983cary} as explained before. All the parameters and generators in the new method are determined during the cancellation process, as explained by the example given in Sec.(\ref{sec1.5}) to reduce the gyrophase from the motion equations of charged particles in magnetized plasmas.

The rest of the paper is arranged as follows. In Sec.(\ref{sec0}), several rules of Lie derivative and pullback transform of differential forms are introduced. In Sec.(\ref{sec1}), a multi-parameter LTPT is derived by extending the single-parameter LTPT. In Sec.(\ref{sec3}), the multi-parameter LTPT given in Sec.(\ref{sec1}) is generalized to a new one by relaxing the limitation over the parameter sequence of the formula of the new 1-form. The way of application and how to derive the coordinate transform are also introduced. As an example of the application, the new electrostatic gyrokinetic model is derived using the multi-parameter Lie transform method in Sec.(\ref{sec1.5}).

\section{A simple introduction to some rules of Lie derivative and pullback transform of the differential 1-form}\label{sec0}

This paper utilizes several rules of Lie derivative and pullback transform of the differential 1-form, which are introduced in this section. To illustrate the linear property of Lie derivative and pullback transform of differential 1-forms, it should be emphasized that, the repeated indexes don't mean the summation of all indexes in this section. However, the Einstein summation rule will be adopted in the remaining sections.
For a vector field $\mathbf{G} = \left( \mathbf{g}^1,\mathbf{g}^2, \cdots ,\mathbf{g}^p \right)$  with the basis ${{\bf{e}}_i} = \frac{\partial }{{\partial {y^i}}}, i\in \{1,\cdots,p\}$, and differential 1-form $\gamma=\sum\limits_{k = 1}^p {{\gamma _k}\left( {\bf{y}} \right)d{y^k}}$ defined on a p-dimensional manifold $\mathbf{y}$ with its coordinate frame also recorded as $\bf{y}$, the Lie derivative of a single ${{\bf{g}}^i} = {g^i}\frac{\partial }{{\partial {y^i}}}$ acting upon $\gamma$ is
\begin{equation}\label{vp7}
{L_{{{\bf{g}}^i}}}\left( {\sum\limits_{k = 1}^p {{\gamma _k}\left( {\bf{y}} \right)d{y^k}} } \right) = \sum\limits_{k = 1}^p {\left[
{g^i}\left( {{\partial _i}{\gamma _k} - {\partial _k}{\gamma _i}} \right)d{y^k}\\
 + \frac{\partial }{{\partial {y^k}}}\left( {{g^i}{\gamma _i}} \right)dy^k
 \right]},
\end{equation}
where repeated indexes don't denote summation. $\gamma_k$ is called the $k$th component of $\gamma$ throughout this paper. It should be noticed that the Lie derivative of the differential 1-form in Eq.(\ref{vp7}) generates a full differential term as the last term on R.H.S of Eq.(\ref{vp7}). This term doesn't contribute to the dynamics determined by the Lagrangian. This property will be used repeatedly in this paper.

It's easy to check the following properties of the Lie derivative of the differential 1-form
\begin{equation}\label{vp9}
{L_{{{\bf{g}}^i} + {{\bf{g}}^j}}}\left( {{\gamma _k}\left( {\bf{y}} \right)d{y^k}} \right) = \left( {{L_{{{\bf{g}}^i}}} + {L_{{{\bf{g}}^j}}}} \right)\left( {{\gamma _k}\left( {\bf{y}} \right)d{y^k}} \right).
\end{equation}
The Lie derivative of the external or full differential term is
\begin{equation}\label{vp11}
{L_{{{\bf{g}}^i}}}\left( {d\gamma } \right) = d\left( {{L_{{{\bf{g}}^i}}}\gamma } \right),
\end{equation}
which shows that the Lie derivative of a full differential term is still a full differential term. This property will be applied repeatedly in this paper.

As for the pullback transform, it's first assumed that $\phi$ is a coordinate transformation defined as ${\phi }:{\bf{Y}} \to {\bf{y}}$ , where $\mathbf{Y}$ is also a $p$-dimensional manifold. $\gamma \left( {{\bf{y}},\varepsilon } \right) = \sum\limits_{h \ge 0} {{\varepsilon ^h}} \sum\limits_{k = 1}^p {{\gamma _{hk}}\left( {\bf{y}} \right)} d{y^k}$ is a differential 1-form defined on $\bf{y}$ with $\varepsilon$ a small parameter, where the subscripts $h$,$k$ denotes the order and the $k$th component of the 1-form, respectively. $\mathbf{v}_i$ is a tangent vector defined on $\mathbf{Y}$ with $i\in \{1,2,\cdots,p\}$. Here, $\mathbf{v}_i$ is chosen to be $\mathbf{v}_i=\partial/\partial Y_i$. $\phi$ induces a pullback transformation ${\phi^*}$ of $\gamma$ written as $\Gamma=\phi^* \gamma$ defined on $\mathbf{Y}$.  By using the standard mathematical terminology, the pullback transform of $\gamma$ is defined as
\begin{equation}\label{a1}
{\left. {\left( {{\phi ^*}\gamma(\mathbf{y},\varepsilon) } \right)} \right|_{\bf{Y}}}\left( {\frac{\partial }{{\partial {Y^i}}}} \right) = {\left. \gamma(\mathbf{y},\varepsilon)  \right|_{\bf{y}}}\left( {{{\left. {d\phi \left( {\frac{\partial }{{\partial {Y^i}}}} \right)} \right|}_{\bf{y}}}} \right),
\end{equation}
based on the contraction rule between the tangent vector and the cotangent vector \cite{2006marsden, 1989arnoldbook}. $\phi^*$ satisfies the following linear property
\begin{equation}\label{vp30}
{\phi ^*}\left( {{\gamma _{hk}}d{y^k} + {\gamma _{lj}}d{y^j}} \right) = {\phi ^*}\left( {{\gamma _{hk}}d{y^k}} \right) + {\phi ^*}\left( {{\gamma _{lj}}d{y^j}} \right),
\end{equation}
which will be used later in deriving the analytical formula of the new 1-form.
The formula ${d\phi \left( {\frac{\partial }{{\partial {Y^i}}}} \right)}$ is a pushforward transformation of $\partial/\partial Y^i$ with the definition $d{\phi }\left( {\frac{\partial }{{\partial {Y^i}}}} \right) \equiv \sum\limits_{k = 1}^p {\frac{{\partial \phi ^k\left( {\bf{Y}} \right)}}{{\partial {Y^i}}}\frac{\partial }{{\partial {y^k}}}}$. Substituting the pushforward transformation to Eq.(\ref{a1}) and adopting the contraction rule, the $i$th component of new 1-form transformed from ${\varepsilon ^h}\sum\limits_{k = 1}^p {{\gamma _{hk}}\left( {\bf{y}} \right)}dy^k$  is
\begin{equation}\label{a2}
{\varepsilon ^h}{\Gamma _{hi}}({\bf{Y}},\varepsilon ) = {\varepsilon ^h}\sum\limits_{k = 1}^p {{\gamma _{hk}}({\phi }({\bf{Y}}) )\frac{{\partial \phi ^k\left( {\bf{Y}} \right)}}{{\partial {Y^i}}}}.
\end{equation}

\section{The multi-parameter LTPT started from a group of autonomous differential equations}\label{sec1}

This kind of multi-parameter LTPT is a direct generalization of the single-parameter LTPT, which begins with a group of autonomous differential equations describing the new coordinate as functions changing along a group of single-parameter vector fields denoted by those infinitesimal generators. The coordinate transform between old and new coordinate frame can be solved from these differential equations given the value of parameter denoting the amplitude of perturbation. The superiority of such differential equations is that they can lead to an analytical solution of the pullback-transform equation for the differential 1-form. But its defect is that the analytical formula of the new 1-form has a restricted structure of the parameter sequence, which needs additional generalization.

\subsection{Cary and Littlejohn's single-parameter LTPT}\label{sec1.1}

Cary and Littlejohn's single-parameter LTPT with a group of autonomous differential equations is first introduced concisely. $\bf{z}$ and $\bf{Z}$ represent old and new coordinate, respectively. The dimension is $p$. $\varepsilon$ is the parameter representing the amplitude of the perturbations, and it's treated as an independent variable in the differential equations. $\bf{Z}$ is a function of $\bf{z}$ and $\varepsilon$, and it's also recorded as $\mathbf{Z}_f(\mathbf{z},\varepsilon)$ in Ref.(\onlinecite{1983cary}), where the coordinate transform is designed to be given by a group of autonomous differential equations
\begin{equation}\label{af1}
\frac{{d Z_f^i}}{{d \varepsilon }}\left( {{\bf{z}},\varepsilon } \right) = {g^i}\left( {\bf{Z}} \right),
\end{equation}
\begin{equation}\label{af2}
\frac{{d{\bf{z}}}}{{d\varepsilon }} = 0,
\end{equation}
for $i\in\{1,2,\cdots,p\}$.  All of the $g^i$s are normalized. The symbol $``d"$ in Eq.(\ref{af2}) denotes the full derivative and Eq.(\ref{af2}) shows that $\bf{z}$ is independent of $\varepsilon$. Given the initial condition $\mathbf{Z}=\mathbf{z}$ at $\varepsilon=0$, Eqs.(\ref{af1},\ref{af2}) provide the solution of $\mathbf{Z}(\mathbf{z},\varepsilon)$. When $\varepsilon$ takes the value of the amplitude of the perturbation, $\mathbf{Z}(\mathbf{z},\varepsilon)$ gives the new coordinate formally, while the generators $g^i(\mathbf{Z})$ serving as the flow fields. Given the coordinate transform between $\bf{z}$ and $\bf{Z}$ derived from Eqs.(\ref{af1},\ref{af2}), the next step of Cary and Littlejohn's LTPT is to derive the new 1-form transformed from the old one.

$\bf{z}$ can also be written as $\mathbf{z}\equiv \mathbf{Z}_b(\mathbf{Z},\varepsilon)$, that is, $\mathbf{z}$ is a function of $\bf{Z}$ and $\varepsilon$. We symbolically write the coordinate transform as $\psi :{\bf{Z}} \equiv {{\bf{Z}}_f}\left( {{\bf{z}},\varepsilon } \right) \to {\bf{z}} \equiv {{\bf{Z}}_b}\left( {{\bf{Z}},\varepsilon } \right)$. It would be useful to note here that the subscripts ``b" and ``f" denote ``backward" and ``forward".
Eq.(\ref{af2}) is rewritten as
\begin{equation}\label{aaff1}
\frac{{\partial Z_b^i}}{{\partial \varepsilon }} + \frac{{d{Z^k}}}{{d\varepsilon }}\frac{{dZ_b^i}}{{d{Z^k}}} = 0.
\end{equation}
Substituting Eq.(\ref{af1}) into Eq.(\ref{aaff1}), the following  equation is derived
\begin{equation}\label{af80}
\frac{{\partial Z_b^i }}{{\partial \varepsilon }}\left( {{\bf{Z}},\varepsilon } \right) =  - g^k\left( {\bf{Z}} \right)\frac{{\partial Z_b^i }}{{\partial {Z^k}}}\left( {{\bf{Z}},\varepsilon } \right),
\end{equation}
where Einstein summation rule over $k\in\{1,2,\cdots,p\}$ is adopted and will be adopted throughout the remaining paper. The solution of Eq.(\ref{af80}) is
\begin{equation}\label{af81}
{{\bf{Z}}_b}\left( {{\bf{Z}},\varepsilon } \right) = \exp \left( { - \varepsilon {g^k}\left( {\bf{Z}} \right){\partial _{{Z^k}}}} \right){\bf{Z}}.
\end{equation}
It's easy to check this solution.

The pullback transform induced by $\psi$ is written as $\psi^*$. For a differential 1-form $\gamma_0(\mathbf{z})\equiv \gamma_0(\mathbf{Z}_b(\mathbf{Z},\varepsilon))$ defined on old coordinate without explicitly dependence on $\varepsilon$, $\Gamma_0\equiv \psi^* (\gamma_0(\mathbf{z}))$ gives the new differential 1-form in new coordinate frame based on Eq.(\ref{a2}), with the $i$th component being
\begin{equation}\label{af82}
{\Gamma _{0i} }\left( {{\bf{Z}},\varepsilon } \right) = \frac{{\partial Z_b^k}}{{\partial Z_{}^i }}\left( {{\bf{Z}},\varepsilon } \right){\gamma _{0k}}\left( {{{\bf{Z}}_b}\left( {{\bf{Z}},\varepsilon } \right) } \right).
\end{equation}
Just as in the derivation given in Ref.(\onlinecite{1983cary}), the analytical solution of Eq.(\ref{af82}) can be derived as
\begin{equation}\label{af4}
\Gamma_{0i}  = \left[ {\exp \left( {\varepsilon {L_{\mathbf{g}}}} \right)\gamma_0 } \right]_i({\bf{Z}}) + \partial S({\bf{Z}})/\partial Z^i,
\end{equation}
with
\begin{equation}\label{af5}
{\left( {{L_\mathbf{g}}\gamma_0 } \right)_j }({\bf{Z}}) = {g^i}(\mathbf{Z})\left( {{\partial _i}{\gamma _{0j} } - {\partial _j }{\gamma _{0i}}} \right)\left( {\bf{Z}} \right).
\end{equation}
The detailed derivation procedure is skipped here and can be found in Ref.(\onlinecite{1983cary}).
If the 1-form in the old coordinate system $\bf{z}$ depends on the same parameter $\varepsilon$, it can be expanded as
\begin{equation}\label{af6}
\gamma ({\bf{z}},\varepsilon ) = {{\varepsilon ^h}{\gamma _h}\left( {\bf{z}} \right)}.
\end{equation}
According to the linear property of $\psi^*$ explained by Eq.(\ref{vp30}), the new 1-form $\Gamma\equiv {\psi ^*}\left( {\gamma ({\bf{z}},\varepsilon )} \right)$ can be written as the sum of the following components
\begin{equation}\label{af83}
 {\Gamma_{hi} }\left( {{\bf{Z}},\varepsilon } \right) = {\varepsilon ^h}\frac{{\partial Z_b^k}}{{\partial Z_{}^i }}\left( {{\bf{Z}},\varepsilon } \right)\left( {{\gamma _{hk}}({\mathbf{Z}_b(\mathbf{Z},\varepsilon)})} \right)
\end{equation}
with $\Gamma_{hi}$ denoting the $i$th component of the new 1-form transformed from the $O(\varepsilon^h)$ part of the old 1-form.
According to Eqs.(\ref{af82},\ref{af4}), for each $\gamma_h(\bf{z})$, the solution is
\begin{equation}\label{af10}
{\Gamma _{hi}} = \left[ {\exp \left( { - \varepsilon {L_\mathbf{g}}} \right){\gamma _h}} \right]_i \left( {\bf{Z}} \right) + \partial S/\partial Z^i.
\end{equation}
Therefore, the full $\Gamma$ is
\begin{equation}\label{af11}
\Gamma  = {\varepsilon ^h}{\Gamma _h}.
\end{equation}

So far, we have shown the formal derivation of the single-parameter LTPT. Given the coordinate transform derived from a group of single-parameter autonomous differential equations, the analytical formula of a new 1-form given by Eq.(\ref{af11}) is derived for any differential 1-form defined on the old coordinate frame. But it should be remembered that all of $g_i$s are formally given. They need to be solved based on additional requirements. We next extend this method to a multi-parameter case.

\subsection{Extending the single-parameter case to a multi-parameter case}\label{sec1.2}

Real physical systems such as the magnetized plasmas may suffer the multi-parameter perturbations,  which make it necessary to generalize the method in Sec.(\ref{sec1.1}) to a multi-parameter case.

We consider a system which includes  $n$ independent parameters contained by the basic parameter set ${E}_n=\{\varepsilon_1,\cdots,\varepsilon_n\}$. These parameters form a basic parameter vector $\mathbf{E}_n=(\varepsilon_1,\cdots,\varepsilon_n)$.  The associated vector of the generator field is recorded as $\mathbf{g}=(\mathbf{g}_1,\mathbf{g}_2,\cdots,\mathbf{g}_n)$. Each $\mathbf{g}_j$ associated with the parameter $\varepsilon_j$ with $j\in \{1,2,\cdots,n\}$ has $p$ components $(g_j^1,\cdots,g_j^p)$ corresponding to the $p$ dimensions of the coordinate frame.  Then, the new coordinate system $\bf{Z}$ can be written as a function like $\mathbf{Z}=\mathbf{Z}_f(\mathbf{z},{E}_n)$. Similar to Eq.(\ref{af1}), the group of differential equations are
\begin{equation}\label{af12}
\frac{{dZ_f^i ({\bf{z}},{E}_n)}}{{d{\varepsilon _j}}} = g_j^i (\bf{Z}),
\end{equation}
for $i\in \{1,\cdots,p\}$, $j\in\{1,2,\cdots,n\}$. And Eq.(\ref{af2}) still stands.

It must be emphasized here that the parameters in ${E}_n$ are independent from each other. For example, ${E}_n$ doesn't include an integer power $\varepsilon^k_j$ for $\varepsilon_j$, neither includes any product between different parameters like $\varepsilon_1 \varepsilon_2$ for $\varepsilon_1$ and $\varepsilon_2$. Since if they are not independent of each other, it will be proven in Sec.(\ref{sec1.4}) that we can't get an analytical solution of the new 1-form similar to Eq.(\ref{af10}).

Similar to Eq.(\ref{af80}), the group of differential equations for $\mathbf{Z}_b(\mathbf{Z},{E}_n)$ are derived as
\begin{equation}\label{af13}
\frac{{\partial Z_b^i}}{{\partial {\varepsilon _j}}}\left( {{\bf{Z}},{{{E}}_n}} \right) =  - g_j^k\left( {\bf{Z}} \right)\frac{{\partial Z_b^i}}{{\partial {Z^k}}}\left( {{\bf{Z}},{{{E}}_n}} \right),
\end{equation}
for each $i$ and $j$.

For a differential 1-form $\gamma(\mathbf{z},{E}_n)$, the new 1-formula transformed from $\gamma$ induced by the coordinate transform is
\begin{equation}\label{af17}
{\Gamma _i }\left( {{\bf{Z}},{{E}}_n} \right) = \frac{{\partial Z_b^k}}{{\partial Z_{}^i }}\left( {{\bf{Z}},{{E}}_n} \right){\gamma _k}\left( {{{\bf{Z}}_b}\left( {{\bf{Z}},{{E}}_n} \right),{{E}}_n} \right),
\end{equation}
for $i,k\in\{1,2,\cdots,p\}$. $\gamma_k$ denotes the $k$th component of $\gamma$.
Next, we will solve ${{\bf{Z}}_b}\left( {{\bf{Z}},{{E}}_n} \right)$ and $\Gamma \left( {{\bf{Z}},{{E}}_n} \right)$ based on Eq.(\ref{af13}) and (\ref{af17}), respectively.

\subsubsection{Solving ${{\bf{Z}}_b}\left( {{\bf{Z}},{{E}}_n} \right)$}\label{sec1.2.1}

For the single-parameter case, it is easy to check that the solution of Eq.(\ref{af80}) is given by Eq.(\ref{af81}).
However, for the multi-parameter case, it's easy to check that the solution is not like ${\bf{z}} = \exp \left( { - \mathbf{E}_n \cdot{{\bf{g}}^k}{\partial _k}} \right){\bf{Z}}$ anymore, where $\bf{g}$ means the vector $(\mathbf{g}_1,\mathbf{g}_2,\cdots,\mathbf{g}_n)$, and an analytical solution is hard to get. On the contrary, the solution is solved order by order based on expanding the original equations given by Eq.(\ref{af13}).

To solve equations given by Eq.(\ref{af13}), $Z_b^i(\mathbf{Z},{E}_n)$ is firstly expanded as
\begin{equation}\label{af15}
Z_b^i(\mathbf{Z},{E}_n)  = {\varepsilon _1^{{m_1}} \cdots \varepsilon _n^{{m_n}}} Z_{b,{m_1} \cdots {m_n}}^i \left( {\bf{Z}} \right)
\end{equation}
with  $m_j\ge 0$ for $j\in\{1,\cdots,n\}$. $Z_{b,0 \cdots 0}^i \left( {\bf{Z}},{E}_n \right) = {Z^i }$ holds. Substituting Eq.(\ref{af15}) back to Eq.(\ref{af13}), $Z_{b,{m_1} \cdots {m_n}}^i \left( {\bf{Z}} \right)$ can be solved order by order. The zero order equation gives the solution for $Z_{b,0 \cdots m_j=1 \cdots 0}^i \left( {\bf{Z}} \right)$ as
\begin{equation}\label{af16}
Z_{b,0 \cdots ({m_j} = 1) \cdots 0}^i \left( {\bf{Z}} \right) =  - g_j^i,
\end{equation}
where only one $m_j$ equals $1$ and others equal zero. The higher order terms can be solved in the same procedure.

\subsubsection{Solving $\Gamma \left( {{\bf{Z}},{{E}}_n} \right)$ with $\gamma(\bf{z})$ independent of all parameters} \label{sec1.2.2}

Just as the single parameter case given in Sec.(\ref{sec1.1}), for the case of multiple parameters, by applying $\partial/\partial \varepsilon_j, j\in \{1,2,\cdots,n\}$ on both sides of Eq.(\ref{af17}), the result can be rearranged to be
\begin{eqnarray}\label{af18}
\frac{{\partial {\Gamma _i}}}{{\partial {\varepsilon _j}}}\left( {{\bf{Z}},{{{E}}_n}} \right) = - g_j^k \left( {\bf{Z}} \right)\left[ {\frac{{\partial {\Gamma _i}}}{{\partial {Z^k }}}\left( {{\bf{Z}},{{{E}}_n}} \right) - \frac{{\partial {\Gamma _k }}}{{\partial {Z^i}}}\left( {{\bf{Z}},{{{E}}_n}} \right)} \right]
 - \frac{\partial }{{\partial {Z^i}}}\left[ {g_j^k \left( {\bf{Z}} \right){\Gamma _k }\left( {{\bf{Z}},{{{E}}_n}} \right)} \right],
\end{eqnarray}
for $i\in \{1,\cdots,p\},j\in\{1,\cdots,n\}$. $\Gamma_i$ represents the $i$th component of the differential 1-form on the new coordinates. Eq.(\ref{af18}) can be simply and symbolically written as
\begin{equation}\label{af19}
\frac{{\partial {\Gamma _i }}}{{\partial {\varepsilon _j}}} =  - {L_{{\mathbf{g}_j}}}\Gamma_i  - d\left( {{{\bf{g}}_j}\cdot\Gamma } \right).
\end{equation}
With Eq.(\ref{af19}), the following formula for the $i$th component can be derived
\begin{equation}\label{af20}
 \frac{{{\partial ^{{m_1}}}}}{{\partial \varepsilon _{{p_1}}^{{m_1}}}} \cdots \frac{{{\partial ^{{m_k}}}}}{{\partial \varepsilon _{{p_k}}^{{m_k}}}}{\Gamma _i }({\bf{Z}},{{E}}_n)
={\left( { - L_{{\mathbf{g}_{{p_1}}}}^{}} \right)^{{m_1}}} \cdots {\left( { - L_{{\mathbf{g}_{{p_k}}}}^{}} \right)^{{m_k}}}\Gamma_i ({\bf{Z}},{{E}}_n) - \frac{\partial S}{\partial Z^i},
\end{equation}
with $m_j\ge 0$  and $k\ge 0$, and  $\varepsilon_{p_j}\in \{\varepsilon_1,\cdots,\varepsilon_n\}$ for $j\in\{1,\cdots,k\}$. In Eq.(\ref{af20}), neighboring $\varepsilon_{p_j}$s are different. Eq.(\ref{af20}) shows that the position of the partial derivative for different $\varepsilon_{p_j}$s is not commutative or the partial derivatives are non-Abelian. Therefore, the Taylor expansion of $\Gamma_i$ is
\begin{eqnarray}\label{af21}
{\Gamma _i}\left( {{\bf{Z}},{{{E}}_n}} \right) =&& \sum\limits_{k \ge 0} {\frac{1}{{\left( {{m_1} +  \cdots  + {m_k}} \right)!}}\sum\limits_{P({\varepsilon _{{p_1}}}, \cdots ,{\varepsilon _{{p_k}}})} {{{\left( {\begin{array}{*{20}{l}}
{\varepsilon _{{p_1}}^{{m_1}} \cdots \varepsilon _{{p_k}}^{{m_k}}}\\
{ \times \frac{{{\partial ^{{m_1}}}}}{{\partial \varepsilon _{{p_1}}^{{m_1}}}} \cdots \frac{{{\partial ^{{m_k}}}}}{{\partial \varepsilon _{{p_k}}^{{m_k}}}}\gamma }
\end{array}} \right)}_i}({\bf{Z}},{{{E}}_n}{{\left. ) \right|}_{{{{E}}_n} =\{0,\cdots,0\}}}} }  \nonumber  \\
&& + \frac{{\partial S\left( {\bf{Z}} \right)}}{{\partial {Z^i}}},
\end{eqnarray}
where symbol $\sum\limits_{P({\varepsilon _{{p_1}}},{\varepsilon _{{p_2}}}, \cdots ,{\varepsilon _{{p_k}}})} {} $ means summation of all permutations of $\left( {\varepsilon _{{p_1}}^{{m_1}},\varepsilon _{{p_2}}^{{m_2}}, \cdots ,\varepsilon _{{p_k}}^{{m_k}}} \right)$ in the partial derivative.
Based on the fact that $\Gamma(\mathbf{Z},{E}_n)=\gamma(\bf{Z})$ when ${E}_n=\{0,\cdots,0\}$, the limit of ${E}_n=\{0,\cdots,0\}$ at both sides of Eq.(\ref{af20}) can be taken. By substituting the result back into Eq.(\ref{af21}), it's derived that
\begin{eqnarray}\label{af22}
&&{\Gamma _i}\left( {{\bf{Z}},{{{E}}_n}} \right) \nonumber \\
&&= \sum\limits_{k \ge 0} {\frac{1}{{\left( {{m_1} +  \cdots  + {m_k}} \right)!}}\sum\limits_{P({\varepsilon _{{p_1}}}, \cdots ,{\varepsilon _{{p_k}}})} {{{\left( {\begin{array}{*{20}{l}}
{{{\left( { - 1} \right)}^{{m_1} +  \cdots  + {m_k}}}\varepsilon _{{p_1}}^{{m_1}} \cdots \varepsilon _{{p_k}}^{{m_k}}}\\
{ \times L_{{{\bf{g}}_{{p_1}}}}^{{m_1}} \cdots L_{{{\bf{g}}_{{p_k}}}}^{{m_k}}\gamma }
\end{array}} \right)}_i}({\bf{Z}},{{{E}}_n}{{\left. ) \right|}_{{{{E}}_n} = 0}}} }  \nonumber \\
&& + \frac{{\partial S\left( {\bf{Z}} \right)}}{{\partial {Z^i}}}
\end{eqnarray}
which can be written compactly as
\begin{equation}\label{af23}
\Gamma_i \left( {{\bf{Z}},{{E}}_n} \right) = [\exp \left( { - {\bf{E}}_n\cdot{L_{\bf{g}}}} \right)\gamma ]_i({\bf{Z}}) +
\frac{\partial S({\bf{Z}})}{\partial Z^i},
\end{equation}
with the inner product ${{\bf{E}}_n}\cdot{L_{\bf{g}}} \equiv \sum\limits_{j = 1}^n {{\varepsilon _j}{L_{{{\bf{g}}_j}}}} $.

\subsubsection{Solving $\Gamma \left( {{\bf{Z}},{{E}}_n} \right)$ with $\gamma(\mathbf{z},{E}_n)$ depending on ${E}_n$ in the old coordinate system} \label{sec1.2.3}

For the case that $\gamma(\mathbf{z},{E}_n)$ explicitly depends on ${E}_n$, the general form of $\gamma(\mathbf{z},{E}_n)$ can be written as
\begin{equation}\label{af24}
\gamma ({\bf{z}},{{E}}_n) =  {\varepsilon _1^{{m_1}}\cdots \varepsilon _n^{{m_n}}{\gamma _{{m_1} \cdots {m_n}}}} ({\bf{z}}),
\end{equation}
where ${{\gamma _{{m_1} \cdots {m_n}}}\left( {\bf{z}} \right)}$ is normalized and $m_j\ge 0$ for $j\in \{1,\cdots,n\}$ .
Just as the single parameter case, the total new 1-form are the summation of ${\varepsilon _1^{{m_1}} \cdots \varepsilon _n^{{m_n}}} {{\Gamma _{{m_1} \cdots {m_n}}}}(\mathbf{Z},{E}_n)$ with ${{\Gamma _{{m_1} \cdots {m_n}}}}(\mathbf{Z},{E}_n)$ transformed from each ${{\gamma _{{m_1} \cdots {m_n}}}}(\bf{z})$, specifically,
\begin{eqnarray}\label{af25}
{\Gamma _{{m_1} \cdots {m_n}}}\left( {{\bf{Z}},{{E}}_n} \right) =  [\exp \left( { - {\bf{E}}_n\cdot{L_{\bf{g}}}} \right){\gamma _{{m_1} \cdots {m_n}}}]({\bf{Z}})
 + d{S_{{m_1} \cdots {m_n}}}({\bf{Z}}),
\end{eqnarray}
\begin{equation}\label{af26}
\Gamma  = {\varepsilon _1^{{m_1}} \cdots \varepsilon _n^{{m_n}}{\Gamma _{{m_1} \cdots {m_n}}}}\left( {{\bf{Z}},{{E}}_n} \right).
\end{equation}

\subsection{ Eq.(\ref{af23}) requires all parameters independent of each other}\label{sec1.4}

To derive the analytical solution in Eq.(\ref{af23}), all parameters need to be independent of each other. To illustrate this, we give a simple counter example which only includes two parameters ${E}_2=\{\varepsilon,\varepsilon^2\}$. The starting two autonomous differential equations are
\begin{equation}\label{af311}
\frac{{dZ_f^i ({\bf{z}},{E}_2)}}{{d\varepsilon }} = g_1^i (\bf{Z}),
\end{equation}
\begin{equation}\label{af31}
\frac{{dZ_f^i ({\bf{z}},{E}_2)}}{{d{\varepsilon ^2}}} = g_2^i (\bf{Z}).
\end{equation}
For a differential 1-form $\gamma(\mathbf{z})$ which isn't explicitly dependent on ${E}_2$ in old coordinates, the two equations induce two differential equations like
\begin{eqnarray}\label{af32}
\frac{{\partial {\Gamma _i}}}{{\partial {{(\varepsilon ^2)}}}}\left( {{\bf{Z}},{{{E}}_2}} \right) =&& - g_2^k \left( {\bf{Z}} \right)\left[ {\frac{{\partial {\Gamma _i}}}{{\partial {Z^k }}}\left( {{\bf{Z}},{{{E}}_2}} \right) - \frac{{\partial {\Gamma _k }}}{{\partial {Z^i}}}\left( {{\bf{Z}},{{{E}}_2}} \right)} \right] \nonumber \\
&&- \frac{\partial }{{\partial {Z^i}}}\left[ {g_2^k \left( {\bf{Z}} \right){\Gamma _k }\left( {{\bf{Z}},{{{E}}_2}} \right)} \right].
\end{eqnarray}
Based on $\Gamma(\mathbf{Z},0)=\gamma(\bf{Z})$, by taking the limit as ${E}_2=\{0,0\}$ in Eq.(\ref{af32}), it's derived that
\begin{eqnarray}\label{af33}
\mathop {\lim }\limits_{{{{E}}_2} \to 0} \frac{{\partial {\Gamma _i}}}{{\partial {(\varepsilon ^2)}}}\left( {{\bf{Z}},{{\bf{E}}_2}} \right) = - g_2^k \left( {\bf{Z}} \right)\left[ {\frac{{\partial {\gamma _i}}}{{\partial {Z^k }}}\left( {\bf{Z}} \right) - \frac{{\partial {\gamma _k }}}{{\partial {Z^i}}}\left( {\bf{Z}} \right)} \right]
 - \frac{\partial }{{\partial {Z^i}}}\left[ {g_2^k \left( {\bf{Z}} \right){\gamma _k }\left( {\bf{Z}} \right)} \right].
\end{eqnarray}
It should be noted here that $\left( {\frac{\partial }{{\partial \varepsilon }}\frac{\partial }{{\partial \varepsilon }}} \right){\Gamma _i }\left( {{\bf{Z}},{{E}}_2} \right) \ne \left( {\frac{{{\partial }}}{{\partial \left( {{\varepsilon ^2}} \right)}}} \right){\Gamma _i }\left( {{\bf{Z}},{{E}}_2} \right)$.

On the other hand, for a 1-form with the expression
\begin{equation}\label{af34}
\Gamma ' = \left[ {\exp \left( { - \varepsilon {L_{{\mathbf{g}_1}}} - {\varepsilon ^2}{L_{{\mathbf{g}_2}}}} \right)\gamma } \right]\left( {\bf{Z}} \right),
\end{equation}
it's derived that
\begin{eqnarray}\label{af35}
\mathop {\lim }\limits_{{{E}}_2 \to \{0,0\}} \frac{{\partial {\Gamma }'}}{{\partial \left( {{\varepsilon ^2}} \right)}}\left( {{\bf{Z}},{{E}}_2} \right) = \frac{1}{2}\left[ {{{\left( {{L_{{\mathbf{g}_1}}}} \right)}^2}\gamma } \right]\left( {\bf{Z}} \right)  - \left( {{L_{{\mathbf{g}_2}}}\gamma } \right)\left( {\bf{Z}} \right),
\end{eqnarray}
which is obviously different from Eq.(\ref{af33}), which doesn't depend on $\mathbf{g}_1$. Therefore, Eq.(\ref{af34}) is not the solution of Eq.(\ref{af33}).

\section{A new method with a given formal analytical formula for the new 1-form  }\label{sec3}

The procedure of the Lie transform method given in Sec.(\ref{sec1}) is as follows. The coordinate transform is first derived from the differential dynamical equations given by Eq.(\ref{af26}). Then, the analytical solution of the new 1-form is derived from the pullback-transform equation Eq.(\ref{af17}), with the result given by Eq.(\ref{af26}). The character of Eq.(\ref{af26}) is that all elements included by the parameter vector $\mathbf{E}_n$ in the exponential formula $\exp \left( { - {{\bf{E}}_n}\cdot{L_{\bf{g}}}} \right)\gamma$ are independent of each other. This character limits its application, as Sec.(\ref{sec1.5}) reveals. For realistic physical problems, the elements included by $\mathbf{E}$ in the operator $\exp \left( { - {\bf{E}}\cdot {L_\mathbf{g}}} \right)$ usually need to include the products between the basic elements like $\{\varepsilon_1,\varepsilon_2,\varepsilon_1 \varepsilon_2\}$. This situation is contradict with Eq.(\ref{af26}). It's necessary to generalize this formalism to a more flexible one.

\subsection{Generalizing the formal formula of the new 1-form}\label{sec3.1}

Here, we presents a generalized version of the muli-parameter LPTP given in Sec.(\ref{sec1}).
Assuming a differential 1-form $\gamma(\mathbf{z},{E}_n)$ given by Eq.(\ref{af24}),
this new method is enlightened by the one given in Sec.(\ref{sec1}) that it formally assumes a solution of the pullback transform of $\gamma(\mathbf{z},{E}_n)$ like
\begin{eqnarray}\label{aff3}
{\Gamma _{{m_1} \cdots {m_n}}}\left( {{\bf{Z}},{{{E}}_n}} \right) =&& [\exp \left( { - {\bf{E}}\cdot{L_{\bf{g}}}} \right){\gamma _{{m_1} \cdots {m_n}}}]({\bf{Z}}) \nonumber \\
&& + d{S_{{m_1} \cdots {m_n}}}({\bf{Z}}),
\end{eqnarray}
with
\begin{equation}\label{aff4}
\Gamma (\mathbf{Z},{E}_n) =  {\varepsilon _1^{{m_1}} \cdots \varepsilon _n^{{m_n}}{\Gamma _{{m_1}\cdots {m_n}}}}(\mathbf{Z},{E}_n).
\end{equation}
The difference between Eq.(\ref{aff3}) and Eq.(\ref{af25}) is that in Eq.(\ref{aff3}), the elements in $\mathbf{E}$ included by the exponential can be any combination of elements in ${E}_n=\{\varepsilon_1,\cdots,\varepsilon_n\}$, e.g, $(\varepsilon_1,\varepsilon_1^2,\varepsilon_1 \varepsilon_2)$.

 So if we also define a group of partial autonomous differential equations like Eq.(\ref{af12}) with the independent variable of each equation being one of the elements in the basic parameters set ${E}_n=\{\varepsilon_1,\cdots,\varepsilon_n\}$, it can be concluded that these equations can not lead to a solution of  $\Gamma  = \left[ {\exp \left( { - {\bf{E}}\cdot{L_{\bf{g}}}} \right)\gamma } \right]\left( {\bf{Z}} \right) + dS\left( {\bf{Z}} \right)$, where $\bf{E}$ includes non-independent parameters as its elements.

\subsection{Deriving the new 1-form by solving $\bf{E}$ and $\bf{g}$}\label{sec3.2}

In the exponential formula in Eq.(\ref{aff3}), $\bf{E}$ and $\bf{g}$ are just formally given. For the application of this method to the practical problems, $\bf{E}$ and $\bf{g}$ are solved based on the following procedure.

First, it needs to formally expand the exponential formula $[\exp \left( { - {\bf{E}}\cdot{L_{\bf{g}}}} \right)\gamma](\bf{Z}) $ order by order like
\begin{equation}\label{aff1}
\Gamma(\mathbf{Z},{E}_n)  = {\varepsilon _1^{{l_1}} \cdots \varepsilon _n^{{l_n}}{\gamma _{{l_1} \cdots {l_n}}}}(\bf{Z}),
\end{equation}
with $l_j$ being an integer for $j\in \{1,\cdots,n\}$ and $\varepsilon_i \in {E}_n$ for all $i\in\{1,\cdots,n\}$. $\gamma_{l_1\cdots l_n}(\mathbf{Z})$ doesn't depend on any parameter.
At each order $O(\varepsilon_1^{l_1} \cdots\varepsilon_n^{l_n})$, those terms depending on fast variables need to be cancelled.

To cancel those perturbative terms depending on fast variables at each order, we  introduce new  and appropriate $\varepsilon_1^{h_1}\cdots \varepsilon_n^{h_n}$ and $\mathbf{g}_{h_1\cdots h_n}$ with $h_i$ being an integer for $i\in\{1,\cdots,n\}$ to the exponential formula to get $\exp \left( { - {\bf{M}} - \varepsilon _1^{{h_1}} \cdots \varepsilon _n^{{h_n}}{L_{{{\bf{g}}_{{h_1} \cdots {h_n}}}}}} \right)$ where $\bf{M}$ denotes the already existed operators. The newly introduced generators generate  new terms depending on fast variables, the lowest order terms of which are designed to cancel the already existed terms depending on fast variables at that order. This cancellation rule will be detailedly explained in Sec.(\ref{sec1.5}) by a specific example.

Therefore, the left terms depending on fast variables possess higher order than that of those cancelled.
By repeatedly introducing new generators for cancellation, the left terms depending on fast variables included by the new 1-form possess higher and higher order, with $\varepsilon_1^{h_1}\cdot\varepsilon_n^{h_n}$s and $\mathbf{g}_{h_1\cdots h_n}$s solved order by order.
For specific Lagrangian 1-form, the complex cancellation calculation can be summarized to be some relatively easy rules. For example, the lowest-order cancellation rule for the Lagrangian 1-form determining the motion of charged particles immersed in strong magnetic field is summarized in Subsec.(\ref{sec1.5.2}). By cancelling out those terms depending on fast variables, the eventual 1-form independent of the fast variables on the new coordinate frame is derived.


\subsection{Solving the coordinate transform}\label{sec3.3}

But it still needs to solve the coordinate transform between the old and the new coordinates. The up procedure doesn't lead to Eqs.(\ref{af12}) and (\ref{af13}) anymore. The equation associating the coordinate transform with the new and the old differential 1-form together is the pullback-transform formula given by Eq.(\ref{af17}).

In Eq.(\ref{af17}),  ${{{\bf{Z}}_b}\left( {{\bf{Z}},{{E}}_n} \right)}$ can be expanded based on basic parameter set ${E}_n=\{\varepsilon_1,\cdots,\varepsilon_n\}$. The expansion of $\mathbf{Z}_b$ is
\begin{equation}\label{af41}
Z_b^i ({\bf{Z}},{{E}}_n) = {Z^\mu } + Z_b^{i *}({\bf{Z}},{{E}}_n),
\end{equation}
and
\begin{equation}\label{af42}
Z_b^{i *}({\bf{Z}},{{E}}_n) = \varepsilon _1^{{m_1}} \cdots \varepsilon _n^{{m_n}}Z_{b,{m_1} \cdots {m_n}}^i ({\bf{Z}}),
\end{equation}
with $m_j\ge 0$ for $j\in \{1,\cdots,n\}$  and the case of all $m_j$s equaling zero deleted.  The formal formula of $\gamma(\mathbf{Z}_b(\mathbf{Z},{E}_n),{E}_n)$ can be written as Eq.(\ref{af24}). Here, $\mathbf{z}\equiv \mathbf{Z}_b(\mathbf{Z},{E}_n)$ is applied. Based on Eq.(\ref{af41}), each ${\gamma _{{m_1} \cdots {m_n}}}\left( {{{\bf{Z}}_b}\left( {{\bf{Z}},{{E}}_n} \right)} \right)$ can be expanded to be
\begin{eqnarray}\label{af43}
&&{\gamma _{{m_1} \cdots {m_n}}}\left( {{{\bf{Z}}_b}\left( {{\bf{Z}},{{E}}_n} \right)} \right) \nonumber \\
&& =  {\frac{{{{\left( {{\bf{Z}}_b^*({\bf{Z}},{{E}}_n)} \right)}^k}}}{{k!}}:{{\left( {\frac{\partial }{{\partial {\bf{Z}}}}} \right)}^k}{\gamma _{{m_1} \cdots {m_n}}}} \left( {\bf{Z}} \right),
\end{eqnarray}
where symbol $:$ means the inner product between two tensors and $k\ge 0$. Assuming we already derived the new 1-form in Subsec.(\ref{sec3.2}), its $i$th component is ${\Gamma _i }\left( {{\bf{Z}},{{E}}_n} \right)$ and can be expanded as
\begin{equation}\label{af44}
{\Gamma _i }({\bf{Z}},{{E}}_n) = \varepsilon _1^{{m_1}} \cdots \varepsilon _n^{{m_n}}\Gamma _{i ,{m_1}\cdots {m_n}}^{}({\bf{Z}}).
\end{equation}
Substituting Eqs.(\ref{af42},\ref{af43},\ref{af44}) back into Eq.(\ref{af17}) and separating the total equation into sub-equations by the order of the elements in basic parameter set ${E}_n$, all $Z_{b,{m_1}{m_2} \cdots {m_n}}^\mu ({\bf{Z}})$s can be solved order by order.

\section{Applying the new multi-parameter Lie transform method to derive the electrostatic gyrokinetic model}\label{sec1.5}

\subsection{The simple introduction of the Lagrangian differential 1-form }\label{sec1.5.1}

To show how to apply this new multi-parameter Lie transform method, we take the example of reducing gyroangle as the fast variable from the Lagrangian differential 1-form, which determines the motion of charged particle in a strong magnetic field with the presentation of the electrostatic perturbation as the mean field. The 1-form is
\begin{equation}\label{af36}
\gamma  = {\bf{A}}\left( {\bf{x}} \right)\cdot d{\bf{x}} + \varepsilon {\bf{v}}\cdot d{\bf{x}} - \left( {\varepsilon \frac{{{{\bf{v}}^2}}}{2} + \phi \left( {{\bf{x}},t} \right)} \right)dt.
\end{equation}
The test particle is chosen from a thermal equilibrium plasma ensemble, e.g., the thermal equilibrium plasma in tokamak. Therefore, $\mathbf{A},\mathbf{v},\mathbf{x},t,\mathbf{B},\mu,\phi$ can be normalized by $A_0=B_0 L_0,v_t,L_0,L_0/v_t,B_0, mv_t^2/B_0,B_0L_0v_t$, respectively. $B_0, L_0$ are characteristic amplitude and spatial length of magnetic field, respectively. $v_t$ is the thermal velocity of particle ensemble the test particle belongs to. And $\varepsilon  \equiv \frac{\rho }{{{L_0}}},\rho \equiv \frac{{m{v_t}}}{{{B_0q}}}$.

The velocity can be written in cylindrical coordinates, by transforming $(\mathbf{x},\mathbf{v})$ to $(\mathbf{x},u_1,\mu_1,\theta)$, where $u_1$ is parallel velocity and $\mu_1$ is magnetic moment. $u_1=\mathbf{v}\cdot \mathbf{b}$ and $\mu_1=v_\perp^2/2B(\mathbf{x})$, and ${\widehat {\bf{v}}_ \bot } = \left( {{{\bf{e}}_1}\sin \theta  + {{\bf{e}}_2}\cos \theta } \right)$. $(\mathbf{e}_1,\mathbf{e}_2,\mathbf{b})$ are orthogonal mutually and $\mathbf{b}$ is the unit vector of the equilibrium magnetic field. After this transformation, $\gamma$ becomes
\begin{eqnarray}\label{a11}
\gamma  = \left( {{\bf{A}}\left( {\bf{x}} \right) + \varepsilon {u_1}{\bf{b}} + \varepsilon \sqrt {2B({\bf{x}}){\mu _1}} {{\widehat {\bf{v}}}_ \bot }} \right)\cdot d{\bf{x}} - \left[ {\varepsilon \left( {\frac{{u_1^2}}{2} + {\mu _1}B({\bf{x}})} \right) + \phi \left( {{\bf{x}},t} \right)} \right]dt
\end{eqnarray}
which can be separated into three parts recorded as
\begin{equation}\label{a12}
{\gamma _0} = {{\bf{A}}}\left( {\bf{x}} \right)\cdot d{\bf{x}},
\end{equation}
\begin{eqnarray}\label{a13}
{\gamma _1} = \varepsilon\left( {{u_1}{\bf{b}} + \sqrt {2B({\bf{x}}){\mu _1}} {{\widehat {\bf{v}}}_ \bot }} \right)\cdot d{\bf{x}} - \varepsilon\left( {\frac{{u_1^2}}{2} + {\mu _1}B({\bf{x}})} \right)dt,
\end{eqnarray}
\begin{equation}\label{a14}
{\gamma _{\sigma } } =  - \phi \left( {{\bf{x}},t} \right)dt.
\end{equation}

$\theta$ is a fast variable and the term depending on $\theta$ in Eq.(\ref{a13}) is of order $O(\varepsilon)$. The $\theta$-dependent term ${\varepsilon }\sqrt {2\mu_1 B({\bf{x}})} {\widehat {\bf{v}}_ \bot }\cdot d{\bf{x}}$ can be treated as the perturbation and $\theta$ can be reduced from the dynamical system up to some order by a coordinate transform as explained by the following content.

The scale of each quantity included in Eq.(\ref{af36}) needs to be explained. In plasma, due to the fact that charged particle can nearly migrate freely in the environment with collective interactions, the magnitude of the potential energy the particles feel must be much smaller than that of its kinetic energy. We define the order of the amplitude of $\phi$ as $O(\varepsilon^\sigma)$ based on the basic parameter $\varepsilon$. The order of the kinetic energy as Eq.(\ref{af36}) shows is $O(\varepsilon)$. Therefore, it's plausible to assume the following range for $\sigma$
\begin{equation}\label{a20}
2 > \sigma>1.
\end{equation}
In Ref.(\onlinecite{1988hahm}), $\sigma$ is chosen to equal $2$. The other quantity needed to mention is the characteristic length scale of the electrostatic potential. Its order is recorded as $O(\varepsilon^{-\beta})$. The case of $0<\beta<1$ is used in the drift kinetic theory\cite{1992hazeltine1}, while the case of $\beta\approx 1$, as the focus of our study, is for the gyrokinetic theory.

\subsection{The lowest-order cancellation rule}\label{sec1.5.2}

Sec.(\ref{sec3}) introduces the general way to reduce the fast variables from the dynamical system. This subsection lists the specific rules for the introduction of new generators into the exponential formula to cancel the $\theta$-dependent terms order by order. According to the mathematical structure of $\gamma$ given in Eq.(\ref{af36}), there are several terms independent of $\theta$ in the Lagrangian 1-form in Eq.(\ref{af36}). Under the new coordinates, these $\theta$-independent terms are combined as $\Upsilon  \equiv {\bf{A}}\cdot d{\bf{X}} + {\Gamma _{1{\bf{X}}\parallel }}\cdot d{\bf{X}} + {\Gamma _{1t}}dt$, with ${\Gamma _{1{\bf{X}}\parallel }} = \varepsilon U{\bf{b}},{\Gamma _{1t}} = \varepsilon (\frac{{{U^2}}}{2} + \mu B({\bf{X}}))$. They are slowly varying terms. The cancellation rule is to introduce new generators, which depend on the gyrophase. The lowest order terms, depending on the gyrophase and generated by the action of the new generators upon those slow-varying terms, are used to cancel the already existing lowest-order $\theta$-dependent terms. According to the results of Lie derivative listed in the appendix, the specific rules are listed below.

(1). Assuming that $[\Gamma_{on\mathbf{X}}]_{\perp} \cdot d\mathbf{X}$ is a 1-form only including $\bf{X}$ component and  $[\Gamma_{on\mathbf{X}}]_\perp$ is a $\theta$-dependent term  perpendicular to the unit vector of magnetic field $\bf{b}$, and its order is $O(\varepsilon^n)$ as the subscript $n$ indicates, we introduce a generator field  $\mathbf{g}_n^{\bf{X}}$ perpendicular to $\mathbf{b}$, to the exponential operator to get $\exp ( - {\varepsilon ^{n }}{L_{\mathbf{g}_{n }^{\mathbf{X}}}} + \mathbf{M})\Upsilon$. $\mathbf{M}$ in the exponential denotes the already existing generators.

The introduction of $\mathbf{g}_n^{\bf{X}}$ produces a linear term $\mathbf{N}_n^{\bf{X}}={\varepsilon ^n}{\bf{g}}_n^{\bf{X}} \times {\bf{B}}\left( {\bf{X}} \right)\cdot d{\bf{X}}$. This term possesses the lowest order among all the generated terms and is used to cancel $[\Gamma_{on \mathbf{X}}]_\perp$.
However, there may also exist other terms possessing the order $O(\varepsilon^n)$ in the $\mathbf{X}$ component and perpendicular to $\mathbf{b}$,  as the result of composite action of several existing generators. These terms also take part in the cancellation, as shown in Eq.(\ref{a26}).

(2). Assuming that $[\Gamma_{on\mathbf{X}}]_{\parallel} \cdot d\mathbf{X}$ is a differential 1-form and  $[\Gamma_{on\mathbf{X}}]_\parallel$ is a $\theta$-dependent term  parallel to $\bf{b}$, and its order is $O(\varepsilon^n)$ as the subscript $n$ indicates, we introduce a generator  $g_{n-1}^U$ to the exponential operator to get $\exp ( - {\varepsilon ^{n - 1}}{L_{g_{n - 1}^U}} +  \cdots )\Upsilon$, of which the lowest-order terms generated is a linear term set
\begin{equation}\label{af88}
{\bf{N}}_{n}^U =  - {\varepsilon ^{n - 1}}{L_{g_{n - 1}^U}}\Upsilon  \\
 =  - {\varepsilon ^n}g_{n - 1}^U{\partial _U}{\Gamma _{1{\bf{X}}\parallel }}\cdot d{\bf{X}} - {\varepsilon ^n}g_{n - 1}^U{\partial _U}{\Gamma _{1t}}dt+d S,
\end{equation}
The first term in ${\bf{N}}_{n}^U$ is used to cancel $[\Gamma_{on,\mathbf{X}}]_\parallel$. The second term is cancelled based on the following point.

(3). Assuming that $\Gamma_{ont}dt $  is a differential 1-form only including $t$ component and it's a $\theta$-dependent term with the order of $O(\varepsilon^n)$, we introduce a generator $g^{\mu}_{n-1}$ to the exponential operator to form $\exp ( - {\varepsilon ^{n - 1}}{L_{g_{n - 1}^\mu }} +  \cdots )\left( {{\Gamma _{1t}}dt} \right)$, of which the lowest-order terms generated are written as
\begin{equation}\label{af89}
{\bf{N}}_n^\mu  =  - {\varepsilon ^{n - 1}}{L_{g_{n - 1}^\mu }}\left( {{\Gamma _{1t}}dt} \right) =  - {\varepsilon ^n}g_{n - 1}^\mu {\partial _\mu }{\gamma _{1t}}dt + dS.
\end{equation}
This term is used to cancel $\Gamma_{ont}$.

Based on points $(1),(2),(3)$, all linear terms in $\mathbf{N}_{n}^{U},\mathbf{N}_{n}^{\mu}$ are completely cancelled. The left $\theta$-dependent terms in the expansion are of order higher than $O(\varepsilon^{n})$. By repeatedly carrying out the lowest-order cancellation, the order of left $\theta$-dependent terms become higher and higher.

\subsection{Reduce the gyrophase from Eq.(\ref{af36}) up to the order $O(\varepsilon^\sigma)$, as the approximation linear to the amplitude of the wave}

\subsubsection{Cancelling $\theta$-dependent terms possessing the order $O(\varepsilon)$}

According to the multi-parameter Lie transform method, the formal formula of the new 1-form on new coordinate can be written as $\Gamma  = \exp \left( { - {\bf{E}}\cdot{L_{\bf{g}}}} \right)\gamma \left( {\bf{Z}} \right)$ where $\mathbf{E}$ and $\mathbf{g}$ need to be solved through reducing $\theta$ from $\Gamma$ up to some order.

Among the expansion, the terms independent of the generators are given as follows
\begin{eqnarray}\label{a16}
{{\bf{M}}_0} =&& \left( {{\bf{A}}\left( {\bf{X}} \right) + \varepsilon U{\bf{b}} + \varepsilon \sqrt {2B({\bf{X}})\mu } {{\widehat {\bf{v}}}_ \bot }} \right)\cdot d{\bf{X}}  \nonumber \\
&& - \left[ {\varepsilon \left( {\frac{{{U^2}}}{2} + \mu B({\bf{X}})} \right) + {\phi }({\bf{X}},t)} \right]dt.
\end{eqnarray}
Term $\varepsilon \sqrt {2B({\bf{X}})\mu } {\widehat {\bf{v}}_ \bot }\cdot d{\bf{X}}$ depends on $\theta$ and needs to be cancelled. According to the lowest order cancellation rule, a generator $\mathbf{g}_1^{\mathbf{X}}$ possessing the order $O(\varepsilon)$ needs to be introduced to the exponential as $\exp(-\varepsilon L_{\mathbf{g}_1^{X}})\gamma(\mathbf{Z})$. The lowest order term it generates is
\begin{eqnarray}\label{a17}
{\bf{N}}_1^\mathbf{X} = \varepsilon {\bf{g}}_1^\mathbf{X} \times {\bf{B}}\left( {\bf{X}} \right)\cdot d{\bf{X}},
\end{eqnarray}
which is used to cancel $\varepsilon \sqrt {2B({\bf{X}})\mu } {\widehat {\bf{v}}_ \bot }\cdot d{\bf{X}}$. ${\bf{g}}_1^\mathbf{X}$ is derived as
\begin{equation}\label{a19}
{\mathbf{g}_1^{{\bf{X}}}} =-{\bm{\rho} _0} \equiv -\sqrt {\frac{{2\mu }}{{B\left( {\bf{X}} \right)}}} \left( { - {{\bf{e}}_1}\cos \theta  + {{\bf{e}}_2}\sin \theta } \right).
\end{equation}
Eventually, it's derived that
\begin{equation}\label{a21}
{\Gamma _1} = \varepsilon U{\bf{b}}\cdot d{\bf{X}}.
\end{equation}

\subsubsection{Cancelling $\theta$-dependent terms possessing the order $O(\varepsilon^\sigma)$}

By expanding $\exp(-\varepsilon L_{\mathbf{g}_1^{X}})\gamma(\mathbf{Z})$, $\mathbf{g}_1^{\mathbf{X}}$ generates a series of terms $\frac{1}{{n!}}{\left( { - \varepsilon L_{{\bf{g}}_1^\mathbf{X}}^{}} \right)^n_\sigma}\left( {{\gamma _{\sigma t}}dt} \right) = \frac{{ - {{\left( { - \varepsilon {\bf{g}}_1^\mathbf{X}\cdot\nabla_\sigma } \right)}^n}\phi \left( {{\bf{X}},t} \right)dt}}{{n!}}$ for $n=1,2,3,\cdots$. The operator ${\left( {L_{{\bf{g}}_1^X}} \right)_\sigma }$ is defined to take action as ${\bf{g}}_1^{\bf{X}}\cdot{\nabla _\sigma } $ with $\nabla_{\sigma}$ acting upon $\phi(\mathbf{X},t)$.
The order of each term is $O(\varepsilon^\sigma)$, since $\nabla_\sigma$ provides a factor of order $O(\varepsilon^{-1})$. The term with $n=1$ is $\varepsilon {\bf{g}}_1^\mathbf{X}\cdot\nabla \phi \left( {{\bf{X}},t} \right)dt$, which depends on $\theta$. According to the lowest-order cancellation rule, it  needs to introduce a generator $g^\mu_{\sigma-1}$ and the associated parameter $\varepsilon^{\sigma-1}$ to get the exponential $\exp \left( { - \varepsilon {L_{{\bf{g}}_1^{\bf{X}}}} - {\varepsilon ^{\sigma-1} }{L_{g_\sigma ^\mu }}} \right)\gamma \left( {\bf{Z}} \right)$. ${\bf{N}}_\sigma ^\mu  =  - {\varepsilon ^\sigma }{L_{g_{\sigma  - 1}^\mu }}\left( {{\gamma _{1t}}dt} \right)$ is the lowest-order term in the $t$ component generated by $g_{\sigma-1}^\mu$ and is used to cancel $\varepsilon {\bf{g}}_1^\mathbf{X}\cdot\nabla \phi \left( {{\bf{X}},t} \right)dt$ with the cancellation equation being
\begin{eqnarray}\label{a22}
&& -\varepsilon L_{{\bf{g}}_1^{\mathbf{X}}}^{}\left( {{\gamma _{\sigma t}}dt} \right) - \varepsilon^{\sigma} L_{g_{\sigma  - 1}^\mu }^{}\left( {{\gamma _{1t}}dt} \right) \nonumber \\
&& = \varepsilon {\bf{g}}_1^{\bf{X}}\cdot\nabla \phi dt + \varepsilon^{\sigma} g_{\sigma  - 1}^\mu B\left( {\bf{X}} \right) = 0.
\end{eqnarray}
The solution of $g_{\sigma-1}^\mu$ provided by Eq.(\ref{a22}) is
\begin{equation}\label{a30}
\varepsilon^{\sigma-1} g_{\sigma-1} ^\mu  =  - \frac{{{\bf{g}}_1^{\bf{X}}\cdot\nabla {\phi(\mathbf{X},t) }}}{{ B\left( {\bf{X}} \right)}}.
\end{equation}

The expanding of $\exp \left( { - \varepsilon {L_{{\bf{g}}_1^{\bf{X}}}} - {\varepsilon ^{\sigma-1} }{L_{g_\sigma ^\mu }}} \right)\gamma \left( {\bf{Z}} \right)$ includes other $\theta$-dependent terms possessing the order $O(\varepsilon^\sigma)$.
Based on Eq.(\ref{a22}), some of these terms can be paired to form the following equality for an arbitrary positive $n$
\begin{equation}\label{a23}
\frac{1}{{n!}}\left( { - \varepsilon L_{{\bf{g}}_1^\mathbf{X}}^{}} \right)_\sigma ^{n}\left( {{\gamma _{\sigma t}}dt} \right) + \frac{1}{{n!}}\left( { - \varepsilon L_{{\bf{g}}_1^\mathbf{X}}^{}} \right)_\sigma ^{n - 1}\left( { - {\varepsilon ^\sigma }L_{g_{\sigma-1}^\mu }^{}} \right)\left( {{\gamma _{1t}}dt} \right) = 0.
\end{equation}
To make sure that the order of these terms in Eq.(\ref{a23}) is $O(\varepsilon^{\sigma})$, all $\nabla_\sigma$s act upon $\phi(\mathbf{X},t)$. Take the following simple example as an illustration
\begin{eqnarray}\label{a24}
{\left( {L_{{\bf{g}}_1^{\bf{X}}}^{}} \right)_\sigma }\left( {{\varepsilon ^{\sigma  - 1}}L_{g_{\sigma  - 1}^\mu }^{}} \right)\left( {{\gamma _{1t}}dt} \right) &=& {\left( {L_{{\bf{g}}_1^{\bf{X}}}^{}} \right)_\sigma }\left( {{\bf{g}}_1^{\bf{X}}\cdot\nabla_\sigma } \right)\phi ({\bf{X}},t)dt  \nonumber \\
 &=& {\left( {{\bf{g}}_1^{\bf{X}}\cdot\nabla_\sigma } \right)^2}\phi ({\bf{X}},t)dt.
\end{eqnarray}

The term $\frac{1}{{n!}}\left( { - \varepsilon L_{{\bf{g}}_1^{\bf{X}}}^{}} \right)_\sigma ^{n - 1}\left( {{\gamma _{\sigma t}}dt} \right)$ for any even positive integer $n$ in Eq.(\ref{a23}) includes the so called Finite Larmor Radius (FLR) terms, which are contained by the orbit equation in the conventional gyrokinetic model. However, these terms are perfectly cancelled in this new method.

The remaining $\theta$-dependent terms possessing the order $O(\varepsilon^\sigma)$ and contained by the expanding of $\exp \left( { - \varepsilon {L_{{\bf{g}}_1^{\bf{X}}}} - {\varepsilon ^{\sigma-1} }{L_{g_\sigma ^\mu }}} \right)\gamma \left( {\bf{Z}} \right)$ are summed below
\begin{eqnarray} \label{a25}
{\bf{SUM}}\cdot d{\bf{X}} &\equiv& \sum\limits_{n = 1} {\left( {\begin{array}{*{20}{l}}
{\frac{{{{\left( { - 1} \right)}^n}{\varepsilon ^{n - 1}}{\varepsilon ^{\sigma} }}}{{n!}}{{\left( {L_{{\bf{g}}_1^{\bf{X}}}^{n - 1}} \right)}_\sigma }L_{g_{\sigma  - 1}^\mu }^{}\left( {{\gamma _{1{\bf{X}} \bot }}\cdot d{\bf{X}}} \right)}\\
{ + \frac{{{{\left( { - 1} \right)}^{n + 1}}{\varepsilon ^{n - 1}}{\varepsilon ^{\sigma} }}}{{\left( {n + 1} \right)!}}{{\left( {L_{{\bf{g}}_1^{\bf{X}}}^{n - 1}} \right)}_\sigma }L_{g_{\sigma  - 1}^\mu }^{}L_{{\bf{g}}_1^{\bf{X}}}^{}\left( {{\bf{A}}\cdot d{\bf{X}}} \right)}
\end{array}} \right)}   \nonumber \\
 &=& \sum\limits_{n = 1} {\frac{{ - 1}}{{n!}}\frac{{\left( {n + 2} \right){\varepsilon ^\sigma }}}{{n + 1}}\left( { - \varepsilon L_{{\bf{g}}_1^{\bf{X}}}^{}} \right)_\sigma ^{n - 1}L_{g_{\sigma  - 1}^\mu }^{}\left( {{\gamma _{1{\bf{X}} \bot }}\cdot d{\bf{X}}} \right)} ,
\end{eqnarray}
based on the equality $O(\bm{\rho}_0\cdot\nabla_\sigma)=O(1)$. In the second equality of Eq.(\ref{a25}), the identity $\varepsilon {L_{{\bf{g}}_1^{\bf{X}}}}\left( {{\bf{A}}\cdot d{\bf{X}}} \right) =  - {\gamma _{1{\bf{X}} \bot }}\cdot d{\bf{X}}$ is used with ${\gamma _{1{\bf{X}}\perp }}$ given by Eq.(\ref{a13}).

The sum given by Eq.(\ref{a25}) only includes $\bf{X}$ components and possesses the order $O(\varepsilon^\sigma)$. It can be  proved that Eq.(\ref{a25}) only includes $\mathbf{X}$ component perpendicular to $\bf{b}$. First, the rule of the operator ${\left( {{L_{{\bf{g}}_1^{\bf{X}}}}} \right)_\sigma }(\mathbf{V}(\mathbf{Z})\cdot d\mathbf{X})$ in Eq.(\ref{a25}) is $- {\bf{g}}_1^{\bf{X}} \times {\nabla _ \bot } \times \mathbf{V}(\mathbf{Z})\cdot d\mathbf{X}$ for any vector function $\mathbf{V}(\mathbf{X})$. The vector generated by $\nabla_\sigma$ is perpendicular to $\mathbf{b}$. Second, in Eq.(\ref{a25}), the differential 1-form $g_{\sigma  - 1}^\mu {\partial _\mu }{\gamma _{1{\bf{X}} \bot }}\cdot d{\bf{X}}$ generated by ${L_{g_{\sigma  - 1}^\mu }}\left( {{\gamma _{1{\bf{X}} \bot }}\cdot d{\bf{X}}} \right)$ only includes the perpendicular $\mathbf{X}$ component. Third, $\mathbf{g}_1^{\mathbf{X}}$ is perpendicular to $\bf{b}$. Based on the three facts, ${\left( {{L_{{\bf{g}}_1^{\bf{X}}}}} \right)_\sigma }\left( {g_{\sigma  - 1}^\mu {\partial _\mu }{\gamma _{1{\bf{X}} \bot }}\cdot d{\bf{X}}} \right)$ forms a cross products between vectors like $ \bm{\bot}  \times  \bm{\bot}  \times  \bm{\bot} \cdot d\mathbf{X}$, the result of which is also a 1-form only including the perpendicular $\mathbf{X}$ component. Therefore, all ${\left( {{L_{{\bf{g}}_1^{\bf{X}}}}} \right)_\sigma }$s in Eq.(\ref{a25}) act upon a perpendicular vector, resulting in an perpendicular $\mathbf{X}$ component, eventually.

To cancel the sum in Eq.(\ref{a25}), according to the lowest-order cancellation rule, it only needs to introduce a generator $\mathbf{g}^{\mathbf{X}}_{\sigma}$ perpendicular to $\bf{b}$ and the associated parameter $\varepsilon^\sigma$ to form the exponential $\exp \left( { - \varepsilon {L_{{\bf{g}}_1^{\bf{X}}}} - {\varepsilon ^{\sigma  - 1}}{L_{g_{\sigma-1} ^\mu }} - {\varepsilon ^\sigma }{L_{{\bf{g}}_\sigma ^{\bf{X}}}}} \right)\gamma \left( {\bf{Z}} \right)$.
The lowest order terms generated by $\mathbf{g}^{\mathbf{X}}_{\sigma}$ possess the order $O(\varepsilon^\sigma)$ and are all used to cancel the sum given by Eq.(\ref{a25}) with the cancellation equation being
\begin{equation}\label{a26}
{\bf{SUM}}\cdot d\mathbf{X} + \sum\limits_{n = 1} {\frac{1}{{n!}}\left( { - \varepsilon L_{{\bf{g}}_1^\mathbf{X}}} \right)_\sigma ^{n - 1}\left( { - {\varepsilon ^\sigma }{L_{{\bf{g}}_\sigma ^\mathbf{X}}}} \right)} \left( {{\bf{A}}\cdot d{\bf{X}}} \right) = 0.
\end{equation}
According to the analysis in the previous paragraph, the second term in Eq.(\ref{a26}) produces only a perpendicular $\mathbf{X}$ component. So the solution $\mathbf{g}_\sigma^{\mathbf{X}}$ in Eq.(\ref{a26}) always exists.


So far, all the $\theta$-dependent terms possessing the order $O(\varepsilon^\sigma)$ are all cancelled. The introduced three generators are independent of $\phi(\mathbf{X},t)$ or linear to $\phi(\mathbf{X},t)$. We only carry out the cancellation of $\theta$-dependent terms to the order $O(\varepsilon^\sigma)$, as the approximation linear to the amplitude of the perturbative wave. If the cancellation continues, the generators become nonlinear to $\phi(\mathbf{X},t)$, as a result of which, the Poisson equation becomes nonlinear to $\phi(\mathbf{X},t)$, causing the difficulty to solve it. Eventually, $\Gamma_\sigma$ is derived as
\begin{equation}\label{a48}
\Gamma_\sigma=-\phi(\mathbf{X},t)dt+dS.
\end{equation}


\subsubsection{The orbit equations approximated up to the order $O(\varepsilon^\sigma)$}

In this paper, terms possessing order equaling or higher than $O(\varepsilon^2)$ are ignored, so that $O(\varepsilon^2)$-order terms like $\frac{{{\varepsilon ^2}}}{2}L_{{\bf{g}}_1^{\bf{X}}}^2\left( {{\bf{A}}\cdot d{\bf{X}}} \right)$ and $\varepsilon {L_{{\bf{g}}_1^{\bf{X}}}}\left( {{\gamma _1}} \right)\left( {\bf{Z}} \right)$ are gotten rid of. By combining the remaining terms in $\mathbf{M}_0$, $\Gamma_1$ and $\Gamma_\sigma$, the eventual $\Gamma$ independent of $\theta$ up to the order $O(\varepsilon^\sigma)$ is
\begin{equation}\label{a27}
\Gamma  = \left( {{\bf{A}}\left( {\bf{X}} \right) + \varepsilon U{\bf{b}}} \right)\cdot d{\bf{X}} - \left[ {\varepsilon \left( {\frac{{{U^2}}}{2} + \mu B({\bf{X}})} \right) + \phi ({\bf{X}},t)} \right]dt.
\end{equation}
An obvious difference from the conventional gyrokinetic theory is that FLR terms don't exist in the new differential 1-form. By imposing the variational principle on the 1-form, the orbit equations can be derived as
\begin{equation}\label{a28}
\mathop {\bf{X}}\limits^. {\rm{ = }}\frac{{U{{\bf{B}}^*} - {\bf{b}} \times \nabla \left( {\varepsilon \mu B\left( {\bf{X}} \right) + \phi ({\bf{X}},t)} \right)}}{{{\bf{b}}\cdot{{\bf{B}}^*}}},
\end{equation}
\begin{equation}\label{a29}
\dot U = \frac{{{{\bf{B}}^*}\cdot\nabla \left( {\varepsilon \mu B\left( {\bf{X}} \right) + \phi ({\bf{X}},t)} \right)}}{{\varepsilon {\bf{b}}\cdot{{\bf{B}}^*}}},
\end{equation}
with ${{\bf{B}}^*}\left( {\bf{X}} \right) = {\bf{B}}\left( {\bf{X}} \right) + \varepsilon U\nabla  \times {\bf{b}}$.

\subsection{The new electrostatic gyrokinetic model derived by this new method.}

The coordinate transform happens between the original coordinate $\mathbf{z}\equiv (\mathbf{x},\mu_1,u_1,\theta_1)$ and the gyrocenter coordinate $\mathbf{Z}\equiv (\mathbf{X},\mu, U,\theta)$. According to Sec.(\ref{sec3.3}), the approximation of the coordinate transform can be derived as
\begin{subequations}\label{a31}
\begin{eqnarray}
{\bf{x}} &=& {\bf{X}} - \varepsilon {\bf{g}}_1^{\bf{X}}({\bf{Z}}),\;\;\;\\
{\mu _1} &=& \mu  - {\varepsilon ^{\sigma  - 1}}g_{\sigma  - 1}^\mu ({\bf{Z}}),\\
{u_1} &=& U,\\
{\theta _1} &=& \theta
\end{eqnarray}
\end{subequations}
$\mathbf{g}_{\sigma}^{\mathbf{X}}$ is ignored in the $\mathbf{X}$ coordinate, since this part contributes a higher order part to the transform of the distribution function.

Then, the units of all physical quantities are recovered. By doing so, the orbit equations with units are
\begin{equation}\label{a32}
\mathop {\bf{X}}\limits^. {\rm{ = }}\frac{{U{{\bf{B}}^*} - {\bf{b}} \times \nabla \left( {\mu B\left( {\bf{X}} \right) + \phi ({\bf{X}},t)} \right)}}{{{\bf{b}}\cdot{{\bf{B}}^*}}},
\end{equation}
\begin{equation}\label{a33}
\dot U = \frac{{{{\bf{B}}^*}\cdot\nabla \left( {\mu B\left( {\bf{X}} \right) + q\phi ({\bf{X}},t)} \right)}}{{m{\bf{b}}\cdot{{\bf{B}}^*}}},
\end{equation}
with ${{\bf{B}}^*} = {\bf{B}} + \frac{{mU}}{q}\nabla  \times {\bf{b}}$. The two formulas for $\bm{\rho}_0$ and $g^\mu_{\sigma-1}$ are
$\bm{\rho}_0=\frac{1}{q}\sqrt {\frac{{2m{\mu}}}{{B\left( {\bf{X}} \right)}}} \left( { - {{\bf{e}}_1}\cos \theta  + {{\bf{e}}_2}\sin \theta } \right)$ and
\begin{equation}\label{a35}
\varepsilon^{\sigma-1} g_{\sigma  - 1}^\mu  = \frac{{q\left( {{\bm{\rho} _0}\cdot\nabla } \right)\phi(\mathbf{X},t) }}{{B\left( {\bf{X}} \right)}}.
\end{equation}
Here, the electrostatic potential $\phi(\mathbf{X},t)$ is the potential at the position $\mathbf{X}$.

For a plasma containing more than one species particles,  the Vlasov equation is
\begin{equation}\label{a34}
\left( {\frac{\partial }{{\partial t}} + \frac{{d{\bf{X}}}}{{dt}}\cdot\nabla  + \frac{d}{{dU}}\frac{\partial }{{\partial U}}} \right)F_s\left( {{\bf{Z}},t} \right) = 0.
\end{equation}
The distribution function ${F_s\left( {{\bf{Z}},t} \right)}$ is on gyrocenter coordinates and subscript $s$ denotes the species. The distribution function on particle coordinate is derived by the following transformation
\begin{equation}\label{a35}
f_s\left(\mathbf{z},t \right) = \int \begin{array}{l}
F_s\left( {{\bf{Z}},t} \right)\delta \left( {{\bf{x}} - {\bf{X}} - {\bm{\rho} _0}}(\mathbf{Z}) \right)\delta \left( {{\mu _1} - \mu  + \varepsilon^{\sigma-1} g_{\sigma  - 1}^\mu } \right)\\
\delta \left( {{u_1} - U} \right)\delta \left( {{\theta _1} - \theta } \right){d^3}{\bf{X}}d\mu dU{{d\theta }}.
\end{array}
\end{equation}
The distribution $F(\mathbf{Z},t)$ can be decomposed as an equilibrium Maxwellian distribution and a perturbative one
\begin{equation}\label{a36}
F_s\left( {{\bf{Z}},t} \right) = {F_{s0}}\left( {{\bf{Z}}} \right) + {F_{s1}}\left( {{\bf{Z}},t} \right).
\end{equation}
The equilibrium Maxwellian distribution function is
\begin{equation}\label{a39}
{F_{s0}}\left( {\bf{Z}} \right) \equiv {n_{s0}}\left( {\bf{X}} \right){\left( {\frac{m_s}{{2\pi {T_{s}}\left( {\bf{X}} \right)}}} \right)^{3/2}}\exp \left( {\frac{{ - m_s{U^2} - \mu B\left( {\bf{X}} \right)}}{{2{T_{s}}\left( {\bf{X}} \right)}}} \right).
\end{equation}
Then, expanding the integral in Eq.(\ref{a35}) and ignoring high order terms, the distribution function on particle coordinates can be linearly approximated as
\begin{eqnarray}\label{a37}
{f_s}\left( {{\bf{z}},t} \right) =&& F_{s0}(\mathbf{z}) - \frac{{{q_s}}}{{{T_sB(\mathbf{x})}}}\left( {\exp \left( { - {\bm{\rho} _0}\left( {\bf{z}} \right)\cdot{\nabla _\sigma }} \right) - 1} \right)\phi \left( {{\bf{x}},t} \right){F_{s0}}\left( {\bf{z}} \right)   \nonumber \\
 && + {F_{s1}}\left( {{\bf{x}} - {\bm{\rho} _0}\left( {\bf{z}} \right),{\mu _1},{u_1},{\theta _1},t} \right).
\end{eqnarray}
This approximation is linear to the amplitude of the perturbative wave.
To get Eq.(\ref{a37}), the following formula is used
\begin{eqnarray}\label{a38}
&& - F_{s0}(\mathbf{z})+\int \begin{array}{l}
 {F_{s0}}\left( {\bf{Z}} \right)\delta \left( {{\bf{x}} - {\bf{X}} - {\bm{\rho} _0(\mathbf{z})}} \right)\delta \left( {{\mu _1} - \mu  + \varepsilon^{\sigma-1}g_{\sigma  - 1}^\mu(\mathbf{z}) } \right)\\
\delta \left( {{u_1} - U} \right)\delta \left( {{\theta _1} - \theta } \right){d^3}{\bf{X}}d\mu dUd\theta
\end{array}   \nonumber  \\
= && - F_{s0}(\mathbf{z})+ {F_{s0}}\left( {{\bf{x}} - {\bm{\rho} _0(\mathbf{z})},{\mu _1} + \varepsilon^{\sigma-1}g_{\sigma  - 1}^\mu(\mathbf{z}) ,{u_1},{\theta _1}} \right) \nonumber \\
 \approx && \sum\limits_{n = 0} {\frac{1}{{(n + 1)!}}\left[ {{{\left( { - {\bm{\rho} _0}\left( {\bf{z}} \right)\cdot{\nabla _\sigma }} \right)}^n}{\varepsilon ^{\sigma  - 1}}g_{\sigma  - 1}^\mu } \right]{\partial _\mu }{F_{s0}}\left( {\bf{z}} \right)}  \nonumber \\
= && \frac{{ - {q_s}}}{{{T_s}B\left( {\bf{x}} \right)}}\sum\limits_{n = 0} {\frac{1}{{(n + 1)!}}\left[ {{{\left( { - {\bm{\rho} _0}\left( {\bf{z}} \right)\cdot{\nabla _\sigma }} \right)}^{n + 1}}\phi \left( {{\bf{z}},t} \right)} \right]} {F_{s0}}\left( {\bf{z}} \right)  \nonumber \\
= && \frac{{ - {q_s}}}{{{T_s}B\left( {\bf{x}} \right)}}\left[ {\exp \left( { - {\bm{\rho} _0}\left( {\bf{z}} \right)\cdot\nabla } \right) - 1} \right]\phi \left( {{\bf{z}},t} \right){F_{s0}}\left( {\bf{z}} \right)
\end{eqnarray}
Here, the order produced by $ {{\bm{\rho} _0}\cdot\nabla_\sigma }  $ is $O(1)$.
The density is obtained by integrating $f(\mathbf{z},t)$ out of the velocity space
\begin{eqnarray}\label{a39}
{n_s}\left( {{\bf{x}},t} \right) =&& n_{s0}(\mathbf{x})  - \frac{{{q_s}}}{{{T_s}}}\left[ {\left\langle {\phi \left( {{\bf{x}} + {\bm{\rho} _0}\left( {\bf{z}} \right),t} \right)} \right\rangle  - {n_{s0}}\phi \left( {{\bf{x}},t} \right)} \right]  \nonumber \\
 && + \int {{F_{s1}}\left( {{\bf{x}} - {\bm{\rho} _0}\left( {\bf{z}} \right),{\mu _1},{u_1},{\theta _1},t} \right)\frac{{B\left( {\bf{x}} \right)}}{{{m_s}}}d{u_1}d{\mu _1}d{\theta _1}} ,
\end{eqnarray}
with the definition
\begin{equation}\label{a42}
\left\langle {\phi \left( {{\bf{x}} + {\bm{\rho} _0}\left( {\bf{z}} \right),t} \right)} \right\rangle  \equiv \int {\exp \left( { - {\bm{\rho} _0}\left( {\bf{z}} \right)\cdot{\nabla _\sigma }} \right)\phi \left( {{\bf{x}},t} \right)} {F_{s0}}\left( {\bf{z}} \right)\frac{1}{{{m_s}}}d{u_1}d{\mu _1}d\theta_1.
\end{equation}
Here, $\frac{B(\mathbf{x})}{m_s}$ is the Jacobian determinant between the rectangular coordinates and $\mathbf{z}$, while the Jacobian determinant between $\mathbf{z}$ and $\mathbf{Z}$ is approximated to equal one.
The Poisson equation becomes
\begin{equation}\label{a40}
 - {\nabla ^2}\phi \left( {{\bf{x}},t} \right) = \frac{1}{\epsilon}\sum\limits_s {{q_s}\left[ {\begin{array}{*{20}{l}}
{ - \frac{{{q_s}}}{{{T_s}}}\left[ {\left\langle {\phi \left( {{\bf{x}} + {\bm{\rho} _0}\left( {\bf{z}} \right),t} \right)} \right\rangle  - {n_{s0}}\phi \left( {{\bf{x}},t} \right)} \right]}\\
{ + \int {{F_{s1}}\left( {{\bf{x}} - {\bm{\rho} _0}\left( {\bf{z}} \right),{\mu _1},{u_1},{\theta _1},t} \right)\frac{{B\left( {\bf{x}} \right)}}{{{m_s}}}} d{u_1}d{\mu _1}d{\theta _1}}
\end{array}} \right]}.
\end{equation}
And the equation for the quasi-neutral condition is
\begin{equation}\label{a41}
\sum\limits_s {{q_s}\left[ {\begin{array}{*{20}{l}}
{ - \frac{{{q_s}}}{{{T_s}}}\left[ {\left\langle {\phi \left( {{\bf{x}} + {\bm{\rho} _0}\left( {\bf{z}} \right),t} \right)} \right\rangle  - {n_{s0}}\phi \left( {{\bf{x}},t} \right)} \right]}\\
{ + \int {{F_{s1}}\left( {{\bf{x}} - {\bm{\rho} _0}\left( {\bf{z}} \right),{\mu _1},{u_1},{\theta _1},t} \right)\frac{{B\left( {\bf{x}} \right)}}{{{m_s}}}} d{u_1}d{\mu _1}d{\theta _1}}
\end{array}} \right]}
\end{equation}

We consider a simple plasma which only includes proton and electron. The distribution of electron uses the adiabatic one. Then, the quasi-neutral equation becomes
\begin{equation}\label{a43}
- \frac{{e}}{{{T_i}}}\left[ {\left\langle {\phi \left( {{\bf{x}} + {\bm{\rho} _0}\left( {\bf{z}} \right),t} \right)} \right\rangle  - {n_0}\left( {\bf{x}} \right) \phi \left( {{\bf{x}},t} \right)} \right] + {{ n}_{i1}} - \frac{{e{n_{0}}\left( {\bf{x}} \right)}}{{{T_e}}}\phi \left( {{\bf{x}},t} \right) = 0
\end{equation}
with the density ${n}_{i1}$ defined as follows
\begin{equation}\label{a44}
{n_{i1}} \equiv \int {{F_{s1}}\left( {{\bf{x}} - {\bm{\rho} _0}\left( {\bf{z}} \right),{\mu _1},{u_1},{\theta _1},t} \right)\frac{{B\left( {\bf{x}} \right)}}{{{m_s}}}} d{u_1}d{\mu _1}d{\theta _1}.
\end{equation}

%

\section{Summary and discussion}\label{sec6}

\subsection{The summary}\label{sec6.1}
In this paper, enlightened by the single-parameter LTPT in Ref.(\onlinecite{1983cary}), we developed a multi-parameter LTPT in Sec.(\ref{sec3}). Although the original paper generalizes the single-parameter LTPT to the high order case, that method can only adapt to the simple perturbations with only one perturbative parameter. For complex perturbations, the formal analytical formula of 1-form can't be given exactly like $\Gamma  = {T_n}{T_{n - 1}} \cdots {T_2}{T_1}\gamma ({\bf{Z}}) + dS$ as Cary and Littlejohn did, since in those problems we can't know order sequence like $(\varepsilon,\varepsilon^2,\varepsilon^3,\cdots)$ in advance. So we need a more general formula of the new 1-form which formally provides a versatile parameter sequence.

The advantage of this new method is proved by using it to derive the gyrokinetic model with the presentation of electrostatic perturbations which possess the multi-scale character. It's found that the new gyrokinetic model is different from the conventional one, since all the FLR terms contained by the orbit equations are cancelled in the new model. The essential difference is that the conventional gyrokinetic model adopts a formal formula of the new differential 1-form like $\Gamma  = \exp \left( { - {L_{{{\bf{g}}_2}}}} \right)\exp \left( { - {L_{{{\bf{g}}_1}}}} \right)\gamma \left( {\bf{Z}} \right)$, where $\mathbf{g}_2$ and $\mathbf{g}_1$ are responsible for gyrocenter and guiding-center transform, respectively, while the new model adopts a formal formula $\Gamma  = \exp \left( { - {\bf{E}}\cdot{L_{\bf{g}}}} \right)\gamma \left( {\bf{Z}} \right)$.  The detailed comparison between the two models will be given in the future work.

\subsection{Discussion about another transform method}\label{sec6.3}

At last, we would like to present a simple discussion about another transform method which is sometimes adopted. This method starts with the coordinate transform formula
\begin{equation}\label{af71}
{\bf{z}} = \exp \left( { - {\bf{E}}\cdot{{\bf{g}}^k}\left( {\bf{Z}} \right){\partial _{{Z_k}}}} \right){\bf{Z}},
\end{equation}
where $\bf{E}$ is a parameter vector and $\bf{g}$ is a generator vector. $\mathbf{g}^k$ denotes the $k$th component of all elements contained by $\bf{g}$. The elements in $\mathbf{E}$ can include any combination of the elements in the basic parameters set $\mathbf{E}_n$. $\bf{E}$ and $\bf{g}$ still need to be solved order by order.

The pullback transform of the differential 1-form induced by the coordinate transform of Eq.(\ref{af71}) is given by Eq.(\ref{af17}). It's hard to get an analytical solution of $\Gamma$ from Eq.(\ref{af17}) with the exponential coordinate transform in Eq.(\ref{af71}). Therefore, the solution of $\Gamma$ is derived by Taylor expanding of Eq.(\ref{af17}) order by order based on the basic parameter set. What we are mostly concerned with is the new Lagrangian differential 1-form. The procedure to reduce the gyroangle based on this method shows that it's much more complex than the one in the text. Therefore, it's probably not appropriate for the practical applications.

\section{Acknowledgement}\label{sec7}

This work was completed at Uji Campus of Kyoto University. This work is partially supported by Grants-in-Aid from JSPS (No.25287153 and
26400531).

\appendix

\section{The results of the Lie derivative on $\gamma$ given by Eq.(\ref{a11})}

Several basic results of the Lie derivative of the Lagrangian 1-form given by Eq.(\ref{a11}) are listed below.
\begin{subequations}
\begin{eqnarray}
 {L_{{{\bf{g}}^{\bf{X}}}}}\left( {{\gamma _0}} \right) &=&  - {{\bf{g}}^{\bf{X}}} \times {\bf{B}}\left( {\bf{X}} \right)\cdot d{\bf{X}} + dS\\
 {L_{{{\bf{g}}^{\bf{X}}}}}\left( {{\gamma _1}} \right) &=&  - {{\bf{g}}^{\bf{X}}} \times \nabla  \times {\gamma _{1{\bf{X}}}}\cdot d{\bf{X}} + {{\bf{g}}^{\bf{X}}}\cdot\nabla {\gamma _{1t}}dt + dS\\
 {L_{{{\bf{g}}^{\bf{X}}}}}\left( {{\gamma _\sigma }} \right) &=& -{{\bf{g}}^{\bf{X}}}\cdot\nabla \phi \left( {{\bf{X}},t} \right)dt + dS\\
 {L_{{g^\mu }}}\left( \gamma  \right) &=& {g^\mu }\left( {{\partial _\mu }{\gamma _{1{\bf{X}} \bot }}\cdot d{\bf{X}} - B\left( {\bf{X}} \right)dt} \right) + dS\\
 {L_{{g^U}}}\left( \gamma  \right) &=& {g^U}\left( {{\partial _U}{\gamma _{1{\bf{X}}\parallel }}\cdot d{\bf{X}} - Udt} \right) + dS
\end{eqnarray}
\end{subequations}




%

%

%

%
%

%


\end{document}